\documentclass[a4paper]{article}

\usepackage{icrc2013}
\usepackage{subfigure}

\def\simleq{\; \raise0.3ex\hbox{$<$\kern-0.75em \raise-1.1ex\hbox{$\sim$}}\; }
\def\simgeq{\; \raise0.3ex\hbox{$>$\kern-0.75em \raise-1.1ex\hbox{$\sim$}}\; }

\newcommand{\GeV}{{\rm GeV}}

\newcommand{\TeV}{{\rm TeV}}

\newcommand{\kpc}{{\rm kpc}}

\title{A unified solution to the anisotropy and gradient problems}

\shorttitle{Solution to CR anisotropy and gradient problems}

\authors{D. Gaggero$^{1,2}$, C. Evoli$^{3}$,  D. Grasso$^{4}$, L. Maccione$^{5,6}$  }

\afiliations{
$^1$SISSA, via Bonomea 265, I-34136, Trieste, Italy\\ 
$^2$INFN, sezione di Trieste, via Valerio 2, I-34127, Trieste, Italy\\
$^3$Institut f\"ur Theoretische Physik, Universit\"{a}t Hamburg, Luruper Chaussee 149, D-22761 Hamburg, Germany\\
$^4$INFN, sezione di Pisa, Largo B. Pontecorvo 3, I-56127 Pisa, Italy\\
$^5$Ludwig-Maximilians-Universit\"{a}t, Theresienstra{\ss}e 37, D-80333 M\"{u}nchen, Germany\\
$^6$Max-Planck-Institut f\"{u}r Physik (Werner Heisenberg Institut), F\"{o}hringer Ring 6, D-80805 M\"{u}nchen, Germany
}

\email{daniele.gaggero@sissa.it}

\abstract{The Fermi-LAT collaboration recently confirmed a discrepancy between the observed longitudinal profile of gamma-ray diffuse emission from the Galaxy and that computed with numerical codes assuming that Cosmic Rays (CRs) are produced by Galactic supernova remnants; the accurate Fermi-LAT measurements make this anomaly hardly explainable in terms of conventional diffusion schemes. Moreover, experimental data from both Muon detector and Extensive Air Shower experiments about the large scale dipole anisotropy of CRs can hardly be compatible with model predictions within the framework of conventional isotropic and homogeneous propagation. We argue that, accounting for a well physically motivated correlation between the CR escape time and the spatially dependent magnetic turbulence power, it is possible to solve both problems at the same time in a very natural way. Indeed, by exploiting this correlation, we find propagation models that fit a wide set of CR primary and secondary spectra, and consistently reproduce the CR anisotropy in the energy range $10^2 - 10^4 \GeV$ and the $\gamma$-ray longitude distribution recently measured by Fermi-LAT. 
}


\begin{document}
\maketitle


\section{Introduction}

\subsection{The anisotropy problem}

It is well known that the propagation of charged CRs in the turbulent Galactic Magnetic fields is well described by a diffusion equation. 
The diffusive motion erases almost completely the information about source positions, leaving only a small residual anisotropy that keeps memory of the location of CR accelerators.
 
Current data (see {\it e.g.}~\cite{Guillian:2005wp} and ref.s therein) show the presence of a large scale dipole anisotropy (LSA);
the LSA is O($10^{-3}$) at TeV energies and shows very weak energy dependence above $\sim10~\TeV$, while below few TeV it depends more strongly on energy and drops down to $\sim10^{-4}$ at 100 GeV.
Also several smaller scale anisotropies are observed in different directions. 

Apart the Compton-Getting effect (yielding an energy independent anisotropy at the $\sim 10^{-4}$ level), the origin of the observed anisotropy can be originated by two processes: {\bf 1)} the CR drift due to inhomogeneous source distribution and global leakage from the Galaxy; {\bf 2)} the stochastic effect of local  sources.

While source stochasticity can be invoked to explain the observed behavior of the anisotropy above 10 TeV \cite{Ptuskin:2006}, it cannot reconcile the anisotropy data with the LSA predicted by diffusive models, in particular models with a strong dependence of the diffusion coefficient upon rigidity $\delta \simeq 0.5$ \cite{Blasi:2011fm}; remarkably, those models are favoured by current B/C and antiproton data as recently shown in \cite{DiBernardo:2009ku}. 

We refer to such discrepancy as the {\em CR anisotropy problem}. 

\subsection{The gradient problem}

Non-local observables, like the galactic $\gamma$-ray interstellar emission, can trace the large scale CR spatial distribution, so they may be a powerful tool to understand the origin of the anisotropy problem. 

It has been known since the EGRET era that, if one computes the cosmic ray (CR) Galactocentric radial distribution adopting a source function deduced from pulsar or supernova remnant (SNR) catalogues, the result appears much steeper than the profile inferred from the $\gamma$-ray diffuse emission along the Galactic plane: the latter appears flatter, with a high contribution from large Galactic radii. 

This discrepancy is known as the {\it $\gamma$-ray gradient problem}. 

A sharp rise of the conversion factor between CO emissivity and ${\rm H_2}$ density (the so called ${\rm X_{CO}}$) with the Galactocentric radius was invoked at the time to fix the problem \cite{gradient_problem_2004}: a larger gas density at large radii compensates for the decreasing CR population and is able to explain the $\gamma$-ray flux detected at high Galactic longitudes.
 
Fermi-LAT confirmed the existence of such a problem \cite{Collaboration:2010cm}. Moreover, the high spatial resolution of the LAT permitted to disentangle the emission coming from the interaction of CRs with the molecular gas (whose modelling is strongly affected by the uncertainty on the ${\rm X_{CO}}$) from the emission originated by the interaction of the Galactic CRs with the atomic gas (whose density is better known from its 21 cm radio emission). An analysis based on $\gamma$-ray maps of the third Galactic quadrant \cite{Collaboration:2010cm} pointed out that the $\gamma$-ray emissivity from {\it neutral} gas (tracing the actual CR density) is indeed flatter than the predicted one confirming the gradient problem independently of the  ${\rm X_{CO}}$. This result led the authors of \cite{Collaboration:2010cm} to look for alternative explanations of the problem, e.g. invoking a thick CR diffusion halo or a source term that becomes flatter at large radii.  Both solutions, however, do not appear completely satisfactory; in particular, a smooth source distribution is in contrast with SNR catalogues. 

\section{Anisotropic inhomogeneous diffusion of CRs}
 
Charged CR do not diffuse isotropically in the turbulent component of the GMF: the regular GMF component, which is oriented almost azimuthally along the spiral arms, produces a deviation from isotropy.

The diffusion tensor can be written in terms of two components, $D_\parallel$ and $D_\perp$, that parametrize respectively the diffusion in the directions parallel and perpendicular to the regular GMF.  The relevance of those components depend on the turbulence level. If the turbulence is weak ($\eta \equiv \delta B/B \ll 1$, where $\delta B$ are the fluctuations of the magnetic field) quasi linear theory (QLT) applies and predicts $D_\perp \approx D_\parallel \eta^2$: so perpendicular diffusion is irrelevant in that case but becomes more and more important as turbulence level increases; in the regime of strong turbulence $(\eta \gg 1)$,  QLT does not apply anymore, and it is argued that isotropy should be restored. 

In our Galaxy the environment is quite complicated, regular and turbulent components have similar strengths $(\eta \simeq 1)$, so QLT does not apply and the properties of diffusion should be studied by means of numerical simulations. 

We refer to the results reported in \cite{DeMarco:2007eh}, where propagation of CR protons in simulated realizations of random MFs  with $\eta = 0.5 \div 2$ and in the presence of a large scale azimuthal field was discussed. 
For the random component, a Kolmogorov spectrum was assumed in agreement with observations \cite{Elmegreen:2004wj}. 
The simulations were performed only for high energies (above $10^{15}$ eV) due to CPU time limitations, but -- since the simulated values of $D_\parallel$ and $D_\perp$ decrease slowly and steadily with decreasing energy --  we assume that their ratio can be extrapolated down to the energies considered in this work. We notice that only the simulated value of the $D_\perp/D_\parallel$ ratio is relevant here since the absolute values of the diffusion coefficient components will be fixed against CR data. 

Concerning the energy dependence of the diffusion coefficients, the main result reported in \cite{DeMarco:2007eh} is that -- while $D_\parallel \propto E^{1/3}$ (in agreement with QLT) --  the slope of $D_\perp$ is significantly steeper: $D_\perp \propto E^{0.5 \div 0.6}$. 

Concerning the behaviour with respect to turbulence level, similarly to QLT predictions, the authors of \cite{DeMarco:2007eh} found $D_\parallel$ and $D_\perp$ to behave oppositely with respect to the turbulent power: $D_\perp$ {\it increases} with turbulence, while $D_\parallel$ {\it decreases}.
This is supported by the following intuitive picture. 
In absence of any turbulence, CRs are expected to undergo spiral motion around the regular field lines (so approximately along the $\phi$ direction, in cylindrical coordinates); if the turbulence level increases, CRs are expected to deviate more and more from these regular trajectories and the random walk in the perpendicular direction is expected to become more and more effective; finally, a higher turbulence reflects in a higher perpendicular diffusion coefficient. 

Comparing the time scales for parallel (along the Galactic spiral arms) and perpendicular escape, we obtained that the perpendicular escape is more relevant in the inner Galaxy and also in the near outer regions, unless a very large halo heigth is assumed (see the main paper \cite{MainPaper} for details).


All the above considerations lead to two main consequences:
\begin{itemize}
\item
Under the observationally preferred astrophysical conditions of Kolmogorov turbulence,  the CR escape time is expected to depend on energy as $E^{0.5 - 0.6}$. Noticeably, this is the same dependence which is favored by a combined analysis of CR nuclei and antiproton spectra \cite{Ptuskin:2005ax,DiBernardo:2009ku}.  
\item
In the inner Galaxy at least,  $T_{\rm esc}  \sim  T_\perp = {H}^2/6 D_\perp$ should be anti-correlated to the turbulent power; this quantity should in turn be correlated to the density of CR sources, which we assume inject turbulence in the ISM. So, regions with higher density of CR sources (in particular the so called {\it molecular ring}, located at R = $4 \div 5$ kpc) are expected to show more turbulence and a more efficient perpendicular escape of Galactic CRs. We will show below that this effect, first pointed out in \cite{Evoli:2008dv}, can lead to a natural solution to both the gradient and anisotropy problem.
\end{itemize}

\section{Solution of the CR gradient problem}

We study the effects of our assumption of a spatial correlation between the diffusion coefficient and the source density by solving the diffusion equation with the 2D version of {\tt DRAGON} numerical diffusion code \cite{dragonweb}, which, differently from other numerical and semi-analytical programs,  is designed to account for a spatially dependent diffusion coefficient.  

The code is 2-dimensional ($R, z$) and assumes a purely azimuthal (no arms) structure of the regular GMF. 

Therefore we can only model perpendicular diffusion and the diffusion coefficient is treated as a (position dependent) scalar. Nevertheless, as only the escape time is relevant to determine the CR density, we can account for parallel diffusion along the spiral arms by using an effective diffusion coefficient  $D_{\rm eff}(R) = {\rm max}\left[ D_\perp(R), (H/R_{\rm arm})^2~D_\parallel(R) \right ]$. 

We assume therefore the phenomenological dependence $D_\perp(R) \propto Q(R)^{\tau}$, where $\tau \simgeq 0$ is a free parameter to be fixed against data (simulations do not allow to determine $\tau$ with sufficient accuracy). According to QLT and numerical simulations we assume $D_\parallel$ to have an opposite dependence on the turbulence strength, hence $D_\parallel(R) \propto Q(R)^{- \tau}$. 

We remark that parallel diffusion has almost no effect on the $\gamma$-ray angular distribution and the local CR anisotropy, as it becomes relevant only in the most external regions of the Galaxy. Its presence, however, naturally prevents the escape time from taking unphysical large values at large radii.

Since now a 3D version of {\tt DRAGON} is available \cite{dragon3D}, we plan to perform in the near future a more detailed analysis in which both the parallel and perpendicular diffusion are treated properly and a realistic model for the spiral arm distribution is implemented.

For the source radial distribution we adopt $Q(R) \propto (R/R_\odot)^{0.2}\exp(-1.4(\frac{R - R_\odot}{R_\odot}))$, based on pulsar catalogues \cite{Lorimer:2006qs}.  Using other, observationally determined, distributions would not change our main results. 

\begin{figure}[h]
	\setlength{\unitlength}{1mm}
     \begin{center}
     \subfigure{
      \includegraphics[width=5.5cm]{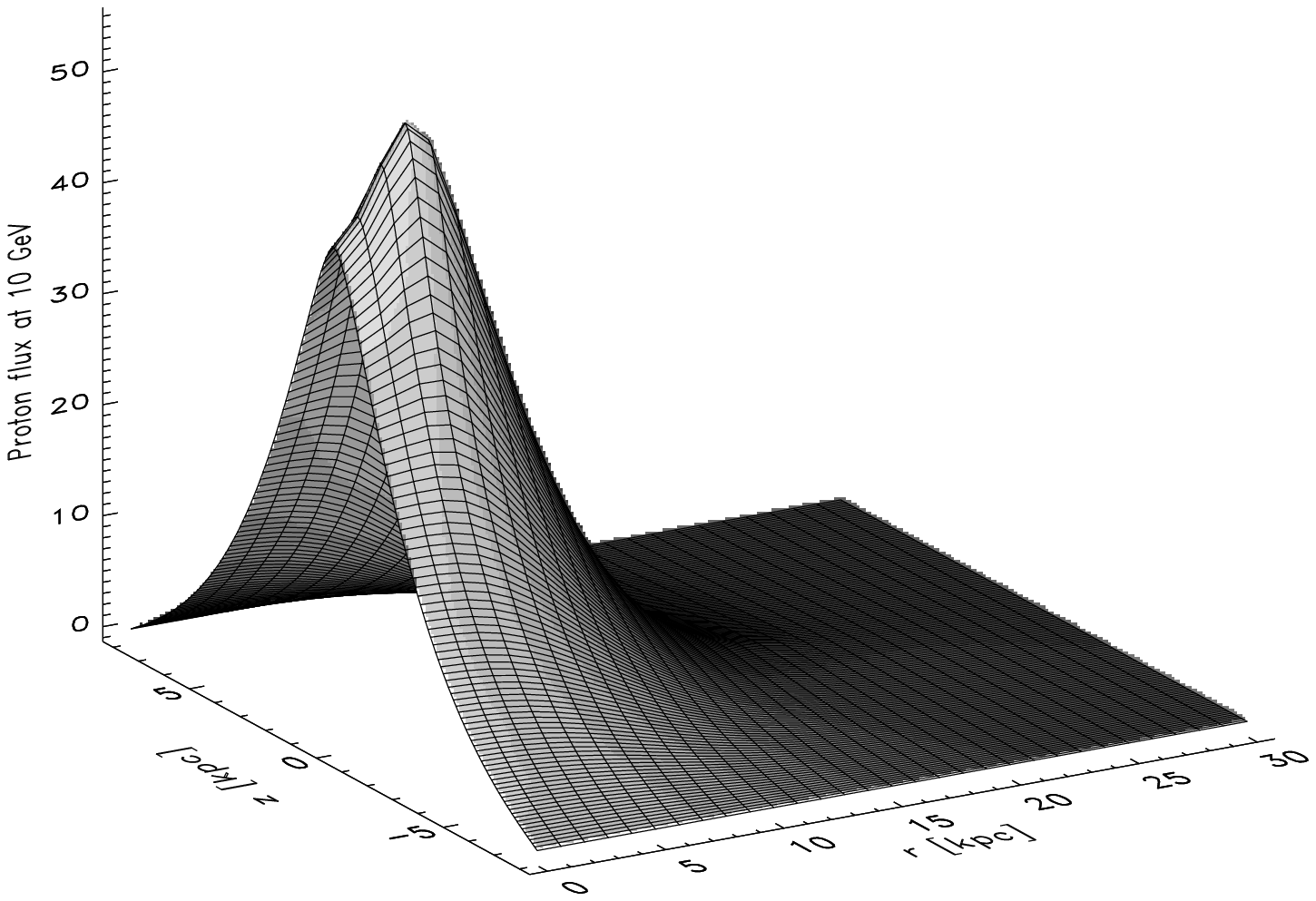}
 	 }
     \subfigure{
      \includegraphics[width=5.5cm]{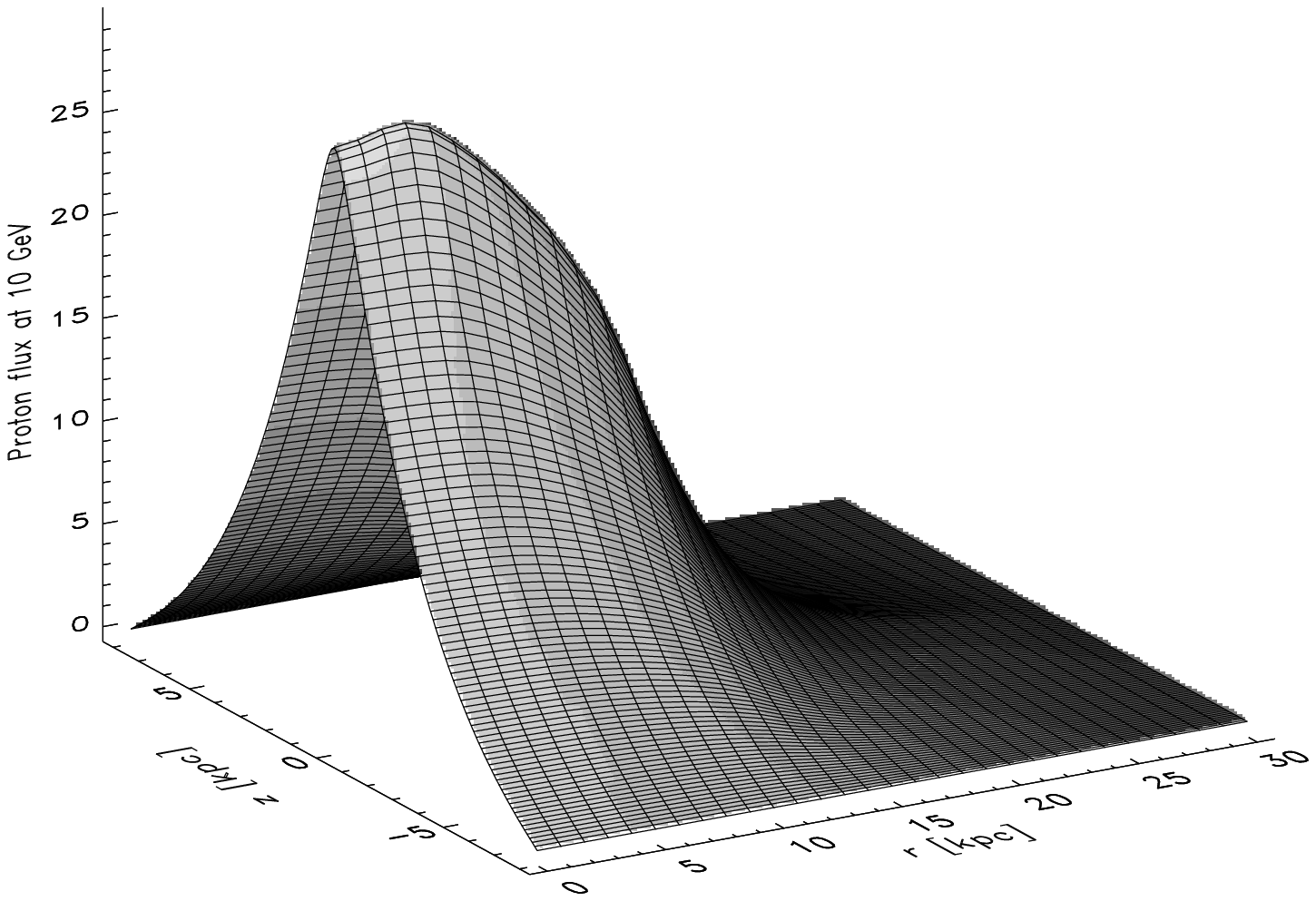}
 	 }
 	 \end{center}
     \caption{\footnotesize \it Two different CR proton distribution maps in arbitrary units computed with {\tt DRAGON} at $10$ GeV are shown as functions of the Galactic cylindrical coordinates $R$ and $z$. {\bf Panel a)} The proton distribution is computed with no radial dependence of diffusion coefficient. {\bf Panel b)} Here the diffusion coefficient is correlated to the source term: $D \propto Q^{\tau}$, with $\tau = 0.8$. The model shows a significant flattening in the CR profile along $R$. The normalization is fixed at $R_{\rm Sun} = 8.5$ kpc in both cases.
}
     \label{fig:protons_map}
\end{figure}

\begin{figure}[tbp]
\centering
\includegraphics[width=0.45\textwidth]{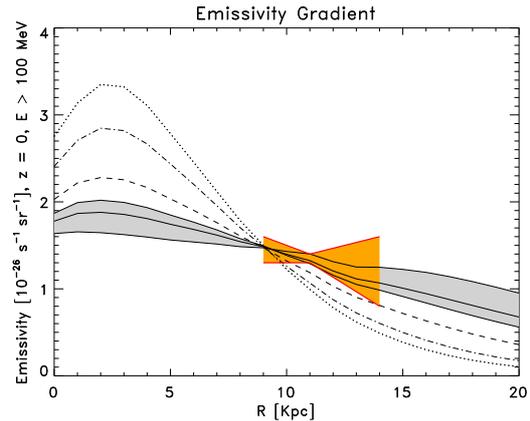}
\caption{\footnotesize \it Integrated $\gamma$-ray emissivity (number of photons emitted per gas atom per unit time) constrained by Fermi-LAT (orange region \cite{Collaboration:2010cm}) compared with our predictions for  $\tau=0,0.2,0.5,0.7,0.8,1.0$ (from top to bottom).
}
\label{fig:PD_exprad_00_02_05_07_08_10_Rmax30}
\end{figure}

Similarly to \cite{DiBernardo:2009ku,Evoli:2008dv} we assume a vertical profile $D_{\rm eff}(R,z) = D_{\rm eff}(R)\exp{(z/H)}$. 
We fix $H = 4~\kpc$ and for each value of $\tau$ we set the $D$ normalization to best fit the B/C and other light nuclei ratios. 
To better highlight the effects of inhomogeneous diffusion we consider here only PD propagation setups. Adding moderate reacceleration does not change significantly any of our results.

We find a good fit of the B/C for all values of $\tau\in[0,1]$. The best fit $D$ normalization only mildly depends on $\tau$. 
Also the computed antiproton and mid-latitude $\gamma$-ray spectra match observations within errors. 

We then calculate the CR density in the Galaxy and the corresponding $\gamma$-ray emissivity (i.e. the number of photons emitted per gas atom per unit time, proportional to CR density) from the CR spatial distributions obtained with our models. 

As clear from Fig.~\ref{fig:PD_exprad_00_02_05_07_08_10_Rmax30}, the model $\tau = 0$ (uniform diffusion) does not reproduce the observed emissivity profile; increasing $\tau$ the gradient of the emissivity profile flattens and for $\tau \simeq {\rm [0.7 \div 0.9]}$ the indirect observation reported in \cite{Collaboration:2010cm} is well reproduced. We also show in Fig. \ref{fig:protons_map} the CR density in the Galaxy as a function of $R$ and $z$ for two cases: $\tau = 0$ and $\tau = 0.8$.

We now calculate the simulated $\gamma$-ray angular distribution by performing a line-of-sight integration of the product of the emissivity times the gas density. For consistency we use the same gas distribution \cite{galpropweb} and the same model for unresolved sources \cite{Strong:2011pa} adopted by the Fermi-LAT collaboration.   

\begin{figure}[h]
	\setlength{\unitlength}{1mm}
     \begin{center}
     \subfigure{
      \includegraphics[width=6.5cm]{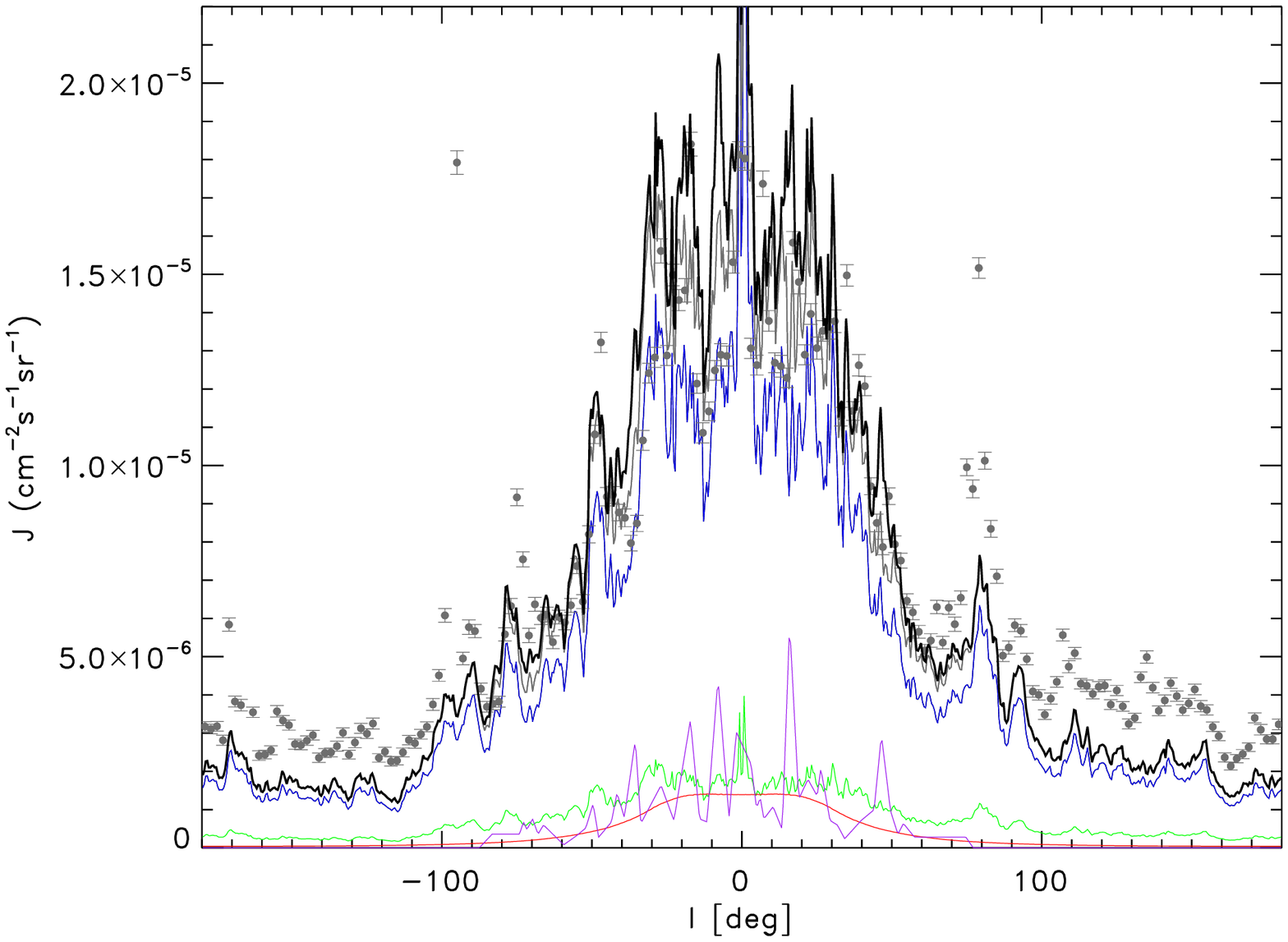}
 	 }
     \subfigure{
      \includegraphics[width=6.5cm]{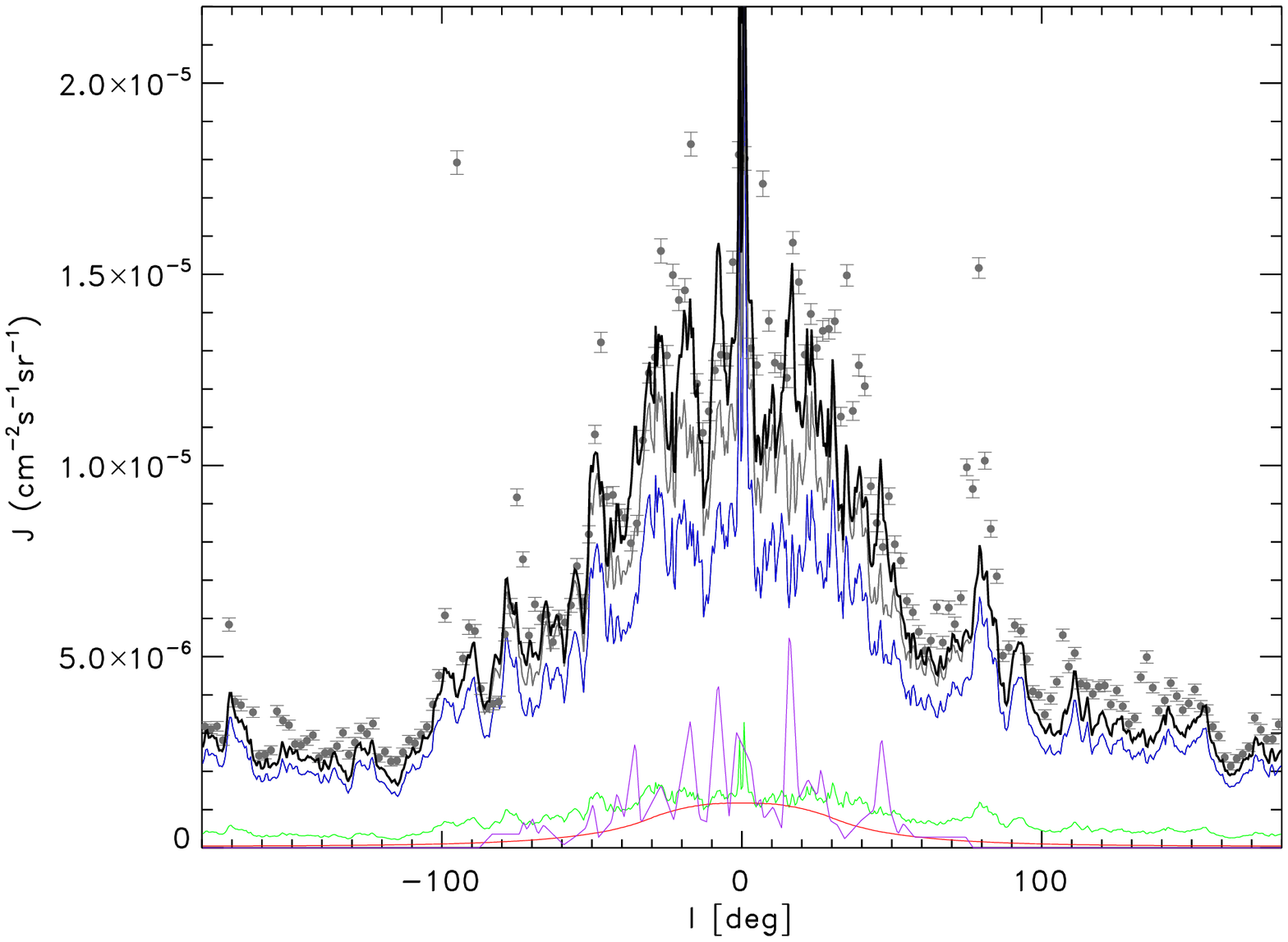}
 	 }
 	 \end{center}
     \caption{\footnotesize \it Two $gamma$-ray longitudinal profiles along the Galactic plane computed with {\tt DRAGON} and {\tt GammaSky} and compared to Fermi-LAT data \cite{Strong:2011pa}. 
Data are integrated over the latitude interval $-5^{\circ} < b < +5^{\circ}$ and in energy between 1104 and 1442 MeV. Red line: IC. Green line: Bremsstrahlung. Blue line: $\pi^0$ decay. Purple line: contribution from unresolved sources. Grey line: $\pi^0$ + IC + Bremsstrahlung. Black line: total. {\bf Panel a)} The profile is computed with no radial dependence of diffusion coefficient. 
     {\bf Panel b)} Here the diffusion coefficient follows the source term: $D \propto Q^{\tau}$, with $\tau = 0.8$. }
     \label{fig:PD_D_exp_exprad}
\end{figure}

We show in Fig. \ref{fig:PD_D_exp_exprad} the longitude profiles of Galactic $\gamma$-ray emission and the residuals of the models against data for $\tau=0$ and $\tau=0.85$. 
The model $\tau = 0$ is clearly too steep compared to the data:  it overshoots the observation in the Galactic center region while it undershoots them by several $\sigma$  in the anti-center region.  
With increasing $\tau$ we get a much smoother behavior of the emissivity in the III quadrant as function of $R$ (see \cite{Collaboration:2010cm} for the possible reasons why the emissivity in the II and III quadrants do not agree entirely). Remarkably, also in thi case, a good match of Fermi-LAT data is achieved for $\tau \simeq {\rm [0.7 \div 0.9]}$, with $\tau = 0.85$ providing an optimal fit and improving the residual distribution.	

\section{Effect on the CR anisotropy}

\begin{figure}[tbp]
\centering
\includegraphics[width=0.45\textwidth]{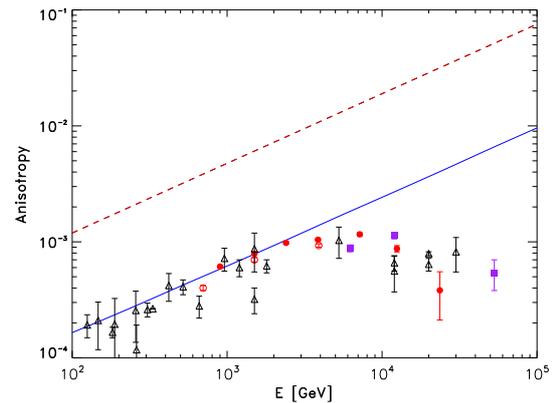}
\caption{\footnotesize \it The CR anisotropy measured by several experiments is compared with our predictions for $\tau=0$ (dashed red line) and $\tau = 0.85$ (solid blue line).}
\label{fig:anisotropy_pd}
\end{figure}

The CR LSA component in the radial direction is related to the CR gradient by
\begin{equation}
{\rm anisotropy} = \frac{3D_{\perp}}{c}  \left\vert{  \frac{\nabla_{r} n_{\rm CR}}{n_{\rm CR}}}\right\vert,
\label{eq:anisotropy}
\end{equation}
which we use to compute the contribution of CR diffusion to the LSA starting from the CR distribution computed in the same PD models as in the previous section. 
Remarkably, with increasing $\tau$, hence with a smoother CR distribution, the predicted LSA also decreases.  Changing from $\tau = 0$ to $\tau = 1$ reduces the anisotropy by almost a factor of 10. 
Intriguingly, we can reproduce the CR anisotropy data \cite{Guillian:2005wp} up to few $\TeV$ with $\tau = 0.85$ (Fig. \ref{fig:anisotropy_pd}). 
The discrepancy between our model and the observed anisotropy above that energy is probably due to the failure of the continuos source distribution approximation we adopted.
Indeed, while below 10 TeV the observed anisotropy phase (see \cite{Guillian:2005wp} and ref.s therein) keeps almost constant to a value compatible with expectations from the global CR leakage, above that energy it significantly fluctuates, as expected if the contribution of stochastic sources becomes dominant.  

 
\section{Conclusions}

We presented a consistent solution to the CR gradient and anisotropy problems. Our approach is based on the physically motivated hypothesis that the CR diffusion coefficient is spatially correlated to the source density: regions in which star, hence SNR, formation is stronger are expected to show a stronger turbulence level and therefore a larger value of the perpendicular diffusion coefficient (oppositely to what happens for $D_\parallel$). The escape of CRs from most active regions is therefore faster, hence smoothing out their density through the Galaxy. Correspondingly, the predicted CR gradient and anisotropy are reduced. We implemented numerically a phenomenological realization of this scenario and checked that, while CR spectral data are still correctly reproduced, our phenomenological approach provides also a remarkably good description of the spectrum and longitude distribution of the diffuse $\gamma$-ray emission measured by the Fermi-LAT collaboration. Our analysis provides for the 
first time a unified propagation model which reproduces local nuclear spectra and also explains non-local observables, and in particular reconciles the preferred low-reacceleration models with $\delta\simeq0.5$ hinted at by the combined spectra of nuclei (B/C), antiprotons, electrons and radio data (and phenomenologically preferred by acceleration theory) with anisotropy and gradient observations. We take these results as an encouragement to pursue a self-consistent theory/computation of non-linear CR - MHD turbulence interaction in the Galaxy.



\vspace{\baselineskip}

\end{document}